\documentclass[prl,twocolumn,showpacs]{revtex4}
\usepackage[dvips]{graphicx}
\usepackage{amsmath,amssymb}
\usepackage{float}
\usepackage{bm}

\newcommand{\ignore}[1]{}
\newcommand{\beq}{\begin{equation}}
\newcommand{\eeq}{\end{equation}}
\newcommand{\mbold}[1]{\mbox{\boldmath $ #1 $}}

\begin{document}

\title
{Quantum Stoner-Wohlfarth model}

\author{Takuya Hatomura$^{1,2}$}  
\email[Corresponding author. Email address: ]{hatomura@spin.phys.s.u-tokyo.ac.jp} 

\author{Bernard Barbara$^{3,4}$}  

\author{Seiji Miyashita$^{1,2}$}

\affiliation
{
$^{1}$ {\it Department of Physics, Graduate School of Science, The University of Tokyo, 7-3-1 Hongo, Bunkyo-Ku, Tokyo, Japan}  \\
$^{2}${\it CREST, JST, 4-1-8 Honcho Kawaguchi, Saitama, 332-0012, Japan}\\
$^{3}${\it Institut N\'eel, CNRS, F-38042, Grenoble, France}\\
$^{4}${\it Universit\'e Grenoble-Alpes/Institut N\'eel, F-38042, Grenoble, France}
}

\date{\today}

\begin{abstract}  
The quantum mechanical counterpart of the famous Stoner-Wohlfarth model  -- an easy-axis magnet in a tilted magnetic field -- is studied theoretically and through simulations, as a function of the spin-size $S$  in a sweeping longitudinal field. Beyond the classical Stoner-Wohlfarth transition, the sweeping field-induced adiabatic change of states slows down as $S$ increases, leading to a dynamical quantum phase transition. This result is described as a critical phenomenon associated with Landau-Zener tunneling gaps at metastable quasi-avoided crossings. Furthermore, a beating of the magnetization is discovered after the Stoner-Wohlfarth transition. The period of the beating, obtained analytically, arises from a new type of quantum phase factor. 
\end{abstract}

\pacs{75.10.Jm, 75.45.+j, 75.50.Xx, 75.60.Jk}

\keywords{ Stoner-Wohlfarth transition, Landau-Zener, non-adiabatic transition}%Use 
\maketitle

We study the reversal of a uniaxial magnet of spin $S$ submitted to a fixed transverse field and a longitudinal sweeping field. This model had been studied for single-molecule magnets with moderate spins $S$=10, in which the Landau-Zener (LZ) transition plays an important role \cite{LZS,BB,Mn12,Fe8,miyaLZS}. Some years ago, the parent model of an Ising spin-chain with ferromagnetic interactions and same longitudinal and transverse fields showed a quantum spinodal phase transition where a size (spin-chain length)-independent magnetization decay was observed and attributed to independent single-spin reversals \cite{QSP}. 

In this letter, we study the quantum aspects of the Stoner-Wohlfarth (SW) transition, and more particularly the relations between the quantum dynamics and classical irreversibility of a large quantum spin $S$ in a uniaxial anisotropy broken by a transverse field, and submitted to a sweeping longitudinal field at zero Kelvin. 
In the $S\rightarrow\infty$ limit, the spin becomes classical and exhibits the so-called SW transition \cite{SW}. This is a 
magnetization jump from a metastable to a stable state when the field, applied in the opposite hemisphere, reaches a critical value. 
We obtained two main results: (i) the classical SW transition (infinite spin) is given by a critical phenomenon of the spinodal type in the limit of a large spin $S$ in the quantum regime, and (ii) when the sweeping field exceeds the SW point, magnetization beatings of quantum mechanical origin, which are analyzed and attributed to a new type of quantum phase factor, appear. 

In order to catch the properties of our model properly in the $S\to\infty$ limit, we introduce the normalized quantum spin operators with a modified commutation relations:
\begin{equation}
s_\alpha=\frac{{S}_\alpha}{S},\ (\alpha=x,y,z), \quad
\left[s_\alpha,s_\beta\right]=\frac{i}{S}\epsilon_{\alpha\beta\gamma}{s}_{\gamma}.
\label{unispin}
\end{equation}
The corresponding SW Hamiltonian, with uniaxial anisotropy, transverse field (fixed) and longitudinal field (sweeping at the time-rate $c$), is written as~\cite{notation}:
\begin{equation}
\mathcal{H}=-D{s}_z^2- H_xs_x-H_zs_z,\ H_z=H_z^{(0)}-ct.
\label{qham}
\end{equation}
In Eq.~(\ref{qham}) and hereafter, we set  $g\mu_{\rm B}=1$ and $\hbar=1$.

The time evolution of the normalized quantum spin operators, given by
\beq
\begin{aligned}
\frac{d{s}_x}{dt}&={1\over S}\left\{D\left({s}_y{s}_z+{s}_z{s}_y\right)+H_z{s}_y\right\}, \\
\frac{d{s}_y}{dt}&={1\over S}\left\{-D\left({s}_z{s}_x+{s}_x{s}_z\right)-H_z{s}_x+H_x{s}_z\right\}, \\
\frac{d{s}_z}{dt}&=-{1\over S}H_x{s}_y, 
\end{aligned}
\label{QSW}
\eeq
is obtained numerically  by the standard Runge-Kutta method.

The corresponding time-evolution of the usual (classical) SW model
comes from the torque equation $d\bm{m}/dt=-\bm{m}\times\bm{H}_\mathrm{eff}$ with the effective field 
$\bm{H}_\mathrm{eff}=-{\partial E_{\rm SW}}/{\partial\bm{m}}=(H_x, 0, 2Dm_z+H_z)$.
The energy of the SW model, $E_{\rm SW}$ , is of course derived from Eq. (\ref{qham}).
Comparing with Eq. (\ref{QSW}), we find that the dynamics becomes the same if we ignore the commutation relations among spin operators in Eq.(\ref{QSW}). But the time in the quantum system should be normalized as
\beq
\tau\equiv {t/S},
\label{scaletime}
\eeq
to study correspondence to the classical dynamics.

Motion of a spin of classical SW model $\bm{m}(t)$ drows a trajectory on the unit sphere. 
If the fields $H_z$ and $H_x$ are weak, due to the anisotropy $D$, the system has metastable state in which the $s_z$ is antiparallel to the direction of $H_z$, but the state is locally stable.
If $H_z$ and $H_x$ increase and the SW condition $(2D)^{2/3} = (H_x)^{2/3} + (H_z)^{2/3}$ is satisfied,
the metastable state becomes unstable. 

\begin{figure}
\centering
$$
\includegraphics[width=4cm]{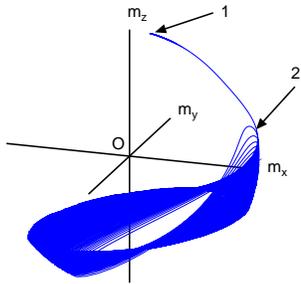}
$$
\caption{ (color online) A classical trajectory under the sweeping field   $H_z=4-ct$ ($c=0.01$) for $D=1$, $H_x=1$. Arrow 1 indicates the starting point which is a metastable fixed point for $H_z=4$, and the arrow 2 the point of the SW jump followed by precession around the magnetic field. }
\label{trD1X1}
\end{figure}

Fig.~\ref{trD1X1} shows the classical motion of magnetization $(m_x(t),m_y(t),m_z(t))$ under a sweeping longitudinal field.  It starts from the metastable point at $H_z(0)=4$ (arrow 1 at $t=0$) and ends at the SW point (arrow 2 $t=t_{\rm SW}$). 
The motion in this process is adiabatic. 
At $t=t_{\rm SW}$ the irreversible magnetization jump takes place. It is followed by a precession-like motion about the local negative effective field, the amplitude of which decreases as $H_z(t)$ becomes larger.

\begin{figure}
$$
\includegraphics[width=7.0cm]{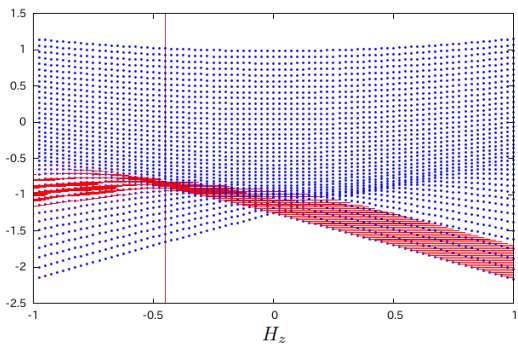}
$$
\caption{(color online) Energy spectrum for $S=20$  with  $D=1$ and $H_x=1$. as a function of $H_z$. For each value of $H_z$ there are $2S+1=41$ eigenvalues which are plotted by blue points.
The brown line shows the position of the SW points.
The red bar denotes the time evolution of population at each eigenstate under a sweeping field $H_z=1-0.08t$. 
}
\label{energylevel}
\end{figure}

\begin{figure}
\centering
\includegraphics[width=7cm]{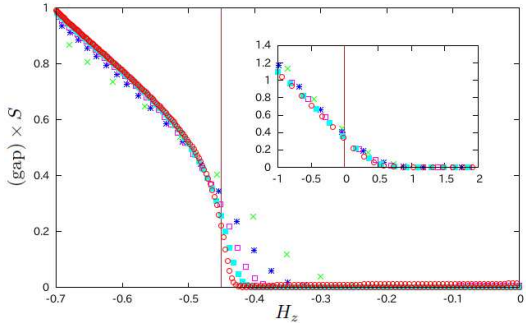}%
\caption{(color online) Normalized energy gaps at avoided-level crossings plotted vs the sweeping magnetic field for $S=20\ (\text{green}),\ 40\ (\text{blue}),\ 80\ (\text{purple}),\ 160\ (\text{cyan})$, and $320\ (\text{red})$ with $H_x=1,\ D=1$.
(inset) Scaling plot showing a mapping of the LZ transition to a dynamical spinodal phase-transition.}
\label{gap}
\end{figure}

To characterize the SW transition of a large quantum spin $S$, we first investigate the energy-level structure as a function of $H_z$. The example $S = 20$ given 
in Fig.~\ref{energylevel} shows how the $2S+1$ levels with positive and negative slopes (i.e. spins), intercept at avoided-level crossings with gaps $\Delta E$.
If a sweeping field is applied to the ground-state (positive side, lowest line), say from $H_z = 1$, this state remains occupied until $H_z = 0$ where the $\pm S$ avoided-level crossing is reached. If the reciprocal sweeping-time is much smaller than $\Delta E$ at this point, the motion is almost adiabatic and the spin-reversal probability is close to unity (LZ model \cite{LZS}). However, in practice this change hardly occurs because, with realistic parameters, 
$\Delta E$ is vanishingly small for a large $S$ and only a very small fraction of the population scatters from $+S$ to $-S$. The rest remains on the line of slope $S$, continuing the $H_z > 0$ ground-state into the $H_z < 0$ region. 
This $H_z < 0$ line corresponds to the classical metastable state and can be called {\it the metastable branch} (See also Fig.~\ref{energylevel}). 
This branch crosses the levels of slopes $M = -S + 1, -S + 2, \cdots, -S+k,\cdots $ successively. At each crossing, some population of the metastable branch scatters to the state of negative magnetization (Fig.~\ref{energylevel}). The Landau-Zener probability of the population remaining at the metastable state, after the $k$-th avoided level-crossing is given by  
\beq
p_{k}^{(S)}=\exp\left(-{\pi\left(\Delta E_k^{(S)}\right)^2\over 4\hbar c^{(S)} \Delta M^{(S)}_k}\right),
\label{pkS}
\eeq 
where $\Delta E_k^{(S)}$ is the energy gap at the field $H_z^k$ and, resulting from the mixing of the spins $S$  (metastable branch) and $-S + k$ (intercepted branches), $\Delta M_k^{(S)}\simeq (2S-k)/S$ is the difference of magnetization (slopes) of the two states.
 As $S$ becomes larger, the spectra associated with a normalized spin densify because the number of the eigenvalues ($2S + $1) increases. For example, going from $S$ to $2S$ leads to a new eigenvalue between two consecutive eigenvalues of the initial spectrum of $S$. 
In the continuous limit, the amount of scattering within a given-field interval must be nearly same, leading to the condition 
$p_{2k}^{(2S)}p_{2k+1}^{(2S)} =p_k^{(S)}$ 
for successive LZ transition probabilities.  
Due to the spin renormalizations ($s_z=S_z/S$, Eq.~(\ref {unispin})),  
one must have 
$\Delta M_{2k}^{(2S)}=\Delta M_k^{(S)}$ and $ct = v\tau$ with  $v = cS$, so that the realization of the probability condition given above implies rescaling the gap, and 
$\Delta E$ becomes $S\Delta E$. 
The latter is plotted vs. $H_z$ (Fig.~\ref{gap}). 

Finite $\Delta E$ means that the magnetization can reverse at fields 
$|H_z| < |H_{\rm SW}|$ with some probabilities, whereas in the $S\to\infty$ limit, the gap is null and the reversal is prohibited as expected in the classical limit. 
This is simply a quantum tunneling effect in the SW model. For example, the quantum SW transition begins to take place at $H_z\sim 2H_{\rm SW}/3$ for $S= 20$ (Fig.~\ref{gap}). 
The dependence of the scaled gap $S\Delta E$ on  the spin changes at
$H_z = H_{\rm SW}$, and it remains finite at $|H_z| >|H_{\rm SW}|$,  which indicates that the spin 
undergos scattering to state of positive magnetization.
This apparently surprising result is simply due to the non-adiabatic character of the SW transition.
The similarity of the field variations of the scaled gaps of Fig.~\ref{gap} suggests some kind of criticality which will now be rapidly investigated.

In the mean-field approach, the spinodal phase transition of a classical system of size $N$, follows the scaling plot $\tau= N^{1/3} f((H-H_{\rm SP})N^{2/3})$, 
where $\tau$ is the relaxation time near the spinodal point 
$H = H_{\rm SP}$ \cite{SPscaling}. 
If our total spin $S$ could be regarded as the size $N$ of a spinodal system, then the data of Fig.~\ref{gap} should be plotted according to the scaling form,
$\tau= S^{1/3} f((H-H_{\rm SP})S^{2/3})$, where $\tau\propto\exp(\pi(SE_k^{(S)})^2/4\hbar c^{(S)}\Delta M_k^{(S)})$. 
In Fig.~\ref{gap}(inset), we find that the data collapse well in a same curve, giving clear evidence of the spinodal character of the quantum SW model. 

\begin{figure}
\includegraphics[width=7cm]{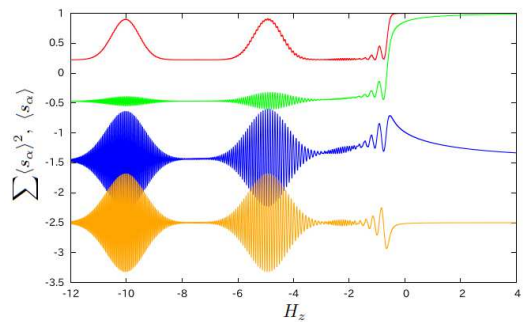}
\caption{
(color online) Beating observed after the SW point. Note that the data are plotted as a function of $H_z$ because $H_z$ is swept linearly in time i.e., $H_z(t) = H_z(0)-ct$.
The red line shows the fidelity $\sum\langle s_\alpha\rangle^2$ and each green, blue, and orange lines show $\langle s_z\rangle$, $\langle s_x\rangle-1.5$, and $\langle s_y\rangle-2.5$ where parameters are $S=20$, $H_x=1$, $D=1$, and $v=0.08$.}
\label{beating}
\end{figure}

We now switch to the last part of this letter, relative to magnetization beatings. When the applied field is swept from right to left in Fig.~\ref{energylevel}, the metastable curve is reached in negative fields inducing population scatterings, across avoided-level crossings, taking place from the SW point. It is in this region that the magnetization beatings are observed (Fig.~\ref{beating}). We define the {\it spin-length fidelity} as $s(t)^2 \equiv \langle s_x(t)\rangle^2+\langle s_y(t)\rangle^2+\langle s_z(t)\rangle^2$.
 This quantity is not conserved, in contrast to the spin modulus $S(S + 1)$. The time evolution of each spin-component and of fidelity is plotted in Fig.~\ref{beating}. 
 The fast oscillations are simple precessions (as in Fig.~\ref{trD1X1}) whereas the beatings are modulations of these precessions. The three magnetization components show the same beating period, which means that the spin-length $s(t)$ is also beating with this period. The amplitude of the $s_z(t)$ beatings decreases as the field becomes large, because large fields induce a tilt of the precession plane towards the $xy$-plane, while the $xy$ components keep the same amplitude. 
 Contrary to precessions, beatings are not seen in the classical case. They can be viewed as a modulation of the precessions of genuine quantum origin. As this will be shown now, they originate from a new type of quantum phase factor. 
The time evolution of the sweeping-field state is given by the time-dependent Schr\"odinger equation $id\Psi/dt={\cal H}\Psi$ where ${\cal H}$ is the scaled Hamiltonian (\ref{qham}) with $H_z^{(0)}= 0$ for simplicity. 
Introducing the unitary transformation to a non-linear acceleration rotation frame
\beq
\psi=U\phi, \quad {\rm with} \quad
U=\exp\left[i\left(D{s}_z^2t-\frac{1}{2}{c}{s}_zt^2\right)\right],
\eeq
it comes
\begin{equation}
i\frac{\partial}{\partial t}\phi=-{H_x} U^\dag{s}_xU\phi\equiv \mathcal{H}_\mathrm{rot}\phi,
\label{hamrot}
\end{equation}
with
\begin{equation}
\mathcal{H}_\mathrm{rot}=
-{H_x}\left\{{s}_x\cos\Theta(\tau)-{s}_y\sin\Theta(\tau)\right\}e^{-iDt/S^2},
\label{Hrot}
\end{equation}
where
\beq
\Theta(\tau)\equiv\frac{1}{2}{v}\tau^2-2D{s}_z\tau.
\label{Theta}
\eeq

\begin{figure}
\includegraphics[width=8cm]{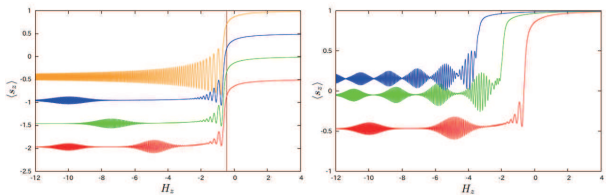}
\caption{
(color online) Dependence of the $z$-component of magnetization dynamics for different values of: (left) spin 
$S=20$ (red), $30$ (green), and 40 (blue) for $v = 0.08$ and $D = 1.0$, %
and (right) anisotropy constant $D=1$ (red), $2$ (green), and 3 (blue) for $S = 20$ and $v = 0.08$. 
The dependence on $v$ ($v=0.08,\ 0.12$, and 0.16 gives the same figure as (left).}
\label{beating2}
\end{figure}

\begin{figure}
$$
\includegraphics[width=8cm]{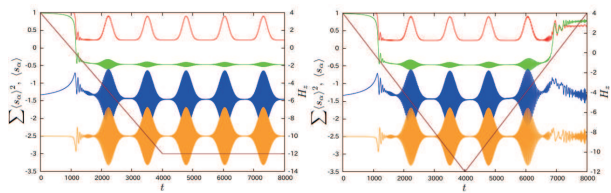}
$$
\caption{
(color online) Beating of the spin-length fidelity, and of the three spin components in a sweeping field above the SW transition for $S = 20$ for $D = 1, H_x = 1$ with the sweeping velocity $c=0.004$ until $H_z(t) = -12$.
 (Left) Beating is not altered after the field (brown curve) is set constant.
 (Right) Beating is suppressed after the field (brown curve) is increased again and 
 re-enters above the $|H_{\rm SW}|$. In these figures, we plot the time evolution as functions of time but not $H_z(t)$ as in the previous figures.
  }
\label{beating3}
\end{figure}

It consists of a Zeeman term in the chosen rotating frame
with classical phase given by expression (\ref{Theta}), multiplied by a quantum phase term $e^{-iDt/S^2}$ (or $e^{-iD\tau/S}$ in $S$ units). 
Note that the sweeping velocity $v=cS =ct/\tau$ intervening in Eq.~(\ref{Hrot}) results from 
Eq.~(\ref{scaletime}). 
The first term, which is independent of $S$, tends to the classical model in the 
$S\rightarrow\infty$ limit. It represents the classical fast precession motion of Fig.~\ref{trD1X1}. The second term $e^{-iDt/S^2}$ depends on $S$ and tends to unity in the $S\rightarrow\infty$ limit. It is of genuine quantum origin and is responsible for the observed beatings. Its specific field-period is given by
\begin{equation}
T_{H_z}={c}T_t={c}\frac{2\pi}{D/S^2}=\frac{2\pi vS}{D},
\end{equation}
where $T_t$ is its specific time-period. The period is proportional to the spin size $S$, the classical sweeping velocity $v$, and inverse proportional to the anisotropy constant $D$, as observed in the simulations (Fig.~\ref{beating2}). 
This confirms that the "beating" phenomenon is due to the factor $e^{-iDt/S^2}$.
Finally, we should note that if the field sweeps back, this quantum beating disappears of course when the field goes through the strongly irreversible SW point. Interestingly, in the reversible region above the SW point, the beatings persist, even if the field is stopped or cancelled at a given time. 
All that is illustrated in Fig.~\ref{beating3} after the third beating, when the field is kept constant (Fig.~\ref{beating3} (left)) or is increased again (Fig.~\ref{beating3} (right)). 

In conclusion, we have studied the dynamics of the classical and the quantum SW models, showing surmountable and impassable bridges between the two and their identification when $S\rightarrow\infty$. 
The two most important results consist in (i) the description of the quantum SW model in terms of a dynamical spinodal phase transition with a scaling of the tunneling gap (or, equivalently, of the time associated with the tunneling probability) vs. $H_z-H_{\rm SW}$.
A detailed description of a population scattering effect (to the states of positive or negative magnetizations) occurring across a succession of LZ transitions along a continuation of the positive-field ground-state in the negative fields above the SW transition (that we called {\it metastable branch}). 
When the field sweeps below the SW transition, the gaps -- and therefore the spin-reversal probabilities -- at avoided-level crossings are so small that the system almost remains metastable, but dynamical transitions to the stable point is still possible. 
However, when $S\rightarrow\infty$ the dynamics disappears leading to the classical, static, SW model at $T=0$. 
A spinodal scaling (Fig.~\ref{gap}) gives a synthetic representation of this complex physics, paving the way for more general studies on the quantum to classical transition. 

This rich physics associated with the metastable branch also leads to our second important results 
(ii) when the sweeping field is larger than the critical SW field, the spin motion, classically described by a time-dependent precession about its slowly moving local-field, becomes modulated in time leading to a beating of the three spin components and of the spin-length fidelity with the characteristic period 
$2\pi vS/D$ obtained analytically thanks to a unitary transformation allowing one to decompose the classical motion and the quantum mechanical phase terms, by which the period of the beating is well explained.
\begin{acknowledgements}
The present work was supported by 
Grants-in-Aid for Scientific Research C (25400391) from MEXT of Japan, and the Elements 
Strategy Initiative Center for Magnetic Materials under the 
outsourcing project of MEXT. The numerical calculations were supported by the supercomputer center of ISSP of University of Tokyo. T.H. is supported by Materials Education program for the future leaders in Research, Industry, and Technology (MERIT). 
\end{acknowledgements}


\begin{thebibliography}{99}
\bibitem{notation}  For the definition of Hamiltonian, we took an convenient form to study the classical limit. 
The present $D$ and $\mbold{H}$ correspond to $DS^2$ and $S\mbold{H}$ of the usual quantum Hamiltonian ${\cal H}=-DS_z^2-\mbold{H}\cdot\mbold{S}$, respectively.
\end{thebibliography}
\end{document}